# Method Cards for Designing for Situational Impairment

**Nayeri Jacobo**
Rochester Institute of Technology
Rochester, NY
nj1919@rit.edu

**ABSTRACT**

Despite significant research that was influenced by Situationally-Induced Impairments and Disabilities (SIIDs) to improve the accessibility of mobile technology, there is still lack of awareness on how to design for SIIDs. Designing for situational impairments does not only affect usability for people who have temporary or long-term disabilities, but also for the "ideal" users who get impacted. Limited resources on how to design for situational impairments overlook inclusive interactions and hinder the creation of accessible technology. Thus, I am going to create method cards that can be used during the design process to figure out how to get designers to design for SIIDs. These method cards help us better understand how to improve the design process by addressing the subject of how to design in order to reduce SIIDs.

**KEYWORDS**

Accessibility; design toolkit; empirical research; situational impairments

**INTRODUCTION**

There is a vast number of people who live with a disability and that number is growing. With more than twenty years of accessible computing research, we are still facing challenges in making mobile technology accessible for people with disabilities [7]. While technology has become fundamental to daily life, several designers and developers overlook the importance of accessibility. There has been limited success in accessible technology that recognizes people with disabilities. Due to this limitation, there is evidence that designers have trouble designing for users that are very unlike themselves and is causing technology not to be adaptive to users' abilities. Not knowing how to properly design for accessibility can hinder the creation of devices to incorporate capabilities to accommodate users' abilities along with environmental factors that can benefit non-disabled and disabled users. Wobbrock recognized the limitation of effective accessible technology that keeps people with disabilities in mind [7]. He coined an approach called Ability-Based Design which focuses on users' abilities and what they can do, rather than centering on what users cannot do [7]. By centering on users' abilities rather than their disability, designers can create interactive systems that are better matched to what users are able to do [7]. After this concept many researchers were able to use the Ability-Based Design approach and created technology that helped both abled and disabled users [2,3,8].

To help design for people with disabilities we need to think about temporary or situational disability. Situational disability occurs when a particular part of a person's environment prevents them from doing a task. When looking at the technologies that were inspired by Ability-Based Design many researchers looked into situational disability as a way to enhance the usability for both disabled and non-disabled users. For example, vision may become sensitive to light after being dilated by a doctor at a routine eye exam. At this current state vision becomes blurry, making interaction with mobile devices difficult, causing situational impairment. When we design with situational disability in mind, the usability of technology improves and it may benefit users with long term disabilities as well as non-disabled users. In this example, situational disability scenarios such as dilated eyes may indirectly lead designers to include solutions that can help visually impaired users. This study intends to further investigate how situational impairments method cards can be used to help designers improve their approach into creating successful design for the creation of accessible technology.

**RELATED WORK**

Despite significant prior research focused on accessibility issues with mobile interactions, there are still countless smartphones that are inaccessible for people with disabilities. Over the years, technology has been embedded into everyday settings, and studies have focused on situations users encounter [4]. Situationally-Induced Impairments and Disabilities, a term coined by Andrew Sears, results from an able-bodied user being affected by both the environment they are working in, and the activity they are engaged in; such that this affects their ability to perform certain tasks [4]. Fifteen years after Sears introduced the concept of situational impairments, researchers began to identify the impact of changing situations in user environments and of the context affecting situational impairments on mobile devices [8]. Studies have also shown that environmental and contextual changes can impact mobile device users in a similar way to how cognitive and physical impairments impact users with disabilities [1]. Contextual factors that can cause users to be situationally impaired range from ambient temperatures, mobile state of the user, physical strain, ambient light, ambient noise, and stress [3]. Research that addresses situational impairments to improve usability and accessibility with mobile devices has shown to address the usability and accessible needs of both people with disabilities and people without.

Users that are situationally impaired by contextual factors when walking might interact with their mobile device similarly to someone who has motor or visual impairments. Research by Yesilda et al. showcased how able-bodied users that were situationally impaired performed a similar number of input errors while typing on a mobile device to a desktop user that has a motor-impairment [8]. Their results had similar conclusions to the research by Barnard et al. Their research is on how environmental and contextual changes impact both abled and disabled users the same way. Using similar techniques to address motor impairments, however, can benefit users with situational impairments by making the device accessible to both kinds of users. As a result, it is best to further explore accessible designs that were influenced by situational impairments to benefit the mobile interaction for users that have disabilities and are abled-bodied.

Research by Goel at el. focused on the lack of investigation on how situational impairment is worsened for mobile text entry on touch screen keyboards. They introduced "WalkType," an adaptive text entry system that controls the built-in accelerometer on a mobile device to counterbalance the extraneous movements while walking [1]. Results from using this tool led to an increase in text entry accuracy while the user is moving. Even though this tool was intended to adapt to someone with a situational impairment, this same tool can benefit someone who is visually impaired or has motor impairments to enhance the accessibility with mobile interaction. Not knowing how to properly design for situational impairments on mobile devices will worsen the performance for visually and motor impaired users [2]. Kane at el. created design guidelines to improve the accessibility on mobile device designs. Within this research, they identified many accessibility issues with mobile devices and addressed how to improve mobile interaction for people with disabilities. Their guidelines suggested incorporating built in sensors for the user's activity and location to allow the device to adapt the interfaces for an increased usability of the device [2]. In this research, Kane et al. demonstrated the importance of including situational impairments during mobile interaction and making mobile interfaces adaptable to contextual changes because it affects the experience for users of all abilities. There have been varied examples on how including situational impairments can encourage more accessible design. Despite significant research, design guidelines, and tools that were influenced by situational disability and were shown to improve accessible technology, there are still many designers who are not implementing accessible design.

In order to bridge this gap, there needs to be more research on how designers use situational impairment in the design process, and how it influences empathy towards disability and accessible design. Research by Shinohara at el. investigated how student designers regard disability and explored how design for multiple disabled and nondisabled users encouraged students to think about accessibility in the design process [5]. When the student designers worked with stakeholders with and without disabilities it helped them understand how impactful accessible design is towards the usability on various users. They were able to gain empathy by interacting with someone with a disability and then creating designs to users' needs. Finally, designers and researchers in the fields of interaction design and HCI have looked into empathic design and have shown the closer designers can get to their users' lives and experience, the more likely that their products could meet user expectations and needs [9].

**METHODOLOGY**

**Plan of Work**
This study intends to understand how to get designers to design for SIIDs because I suspect designers don't know how to design for Situationally-Induced Impairments and Disabilities. I also suspect this is because current technology is unusable in impaired situations or only designed to address one level of SIIDs. Research by Tigwell et al.'s explains that designing for situational visual impairments (SVIs) is understandably 'behind' accessibility design: designing to reduce SVIs is often not in the design scope or part of the designer's current practice [6]. I am going to create method cards to better understand how to improve the design process by addressing how to get designers to reduce SIIDs. I chose method cards because design tools are one method of supporting action in practice, and it was a requested solution from designers to learn how to design to reduce SVIs [6]. If designers are requesting these design tools for SVIs, we can extend the tools to address a variety of SIIDs. The creation of method cards allows us to improve the design process and guide designers towards making informed decisions in order to reduce SIIDs and not dictate a solution. Microsoft Design Toolkit and HaptiMap workbook show awareness about situational impairments, but lack informed guidelines on how to design to reduce SIIDs.

The approach towards designing method cards for SIIDs is to use a user-centered design method with designers. This allows us to discover what information in the tool is needed to help guide designers towards designing for SIIDs. I have the research that showcase the different scenarios that causes SIIDs. I need to properly translate these findings to speak the designer's language and fit their environment. Since there are limited resources on how to help designers design for SIIDs I will seek feedback from this workshop on how to improve the method cards. This contribution can help further discover new approaches towards designing guidelines and tools that can fit the designers needs towards reducing SIIDs.

**CONCLUSIONS**

Situationally-Induced Impairments and Disabilities has unique design properties: they are sometimes accessibility issues and sometimes typical usability issues. My research contributes to the workshop's topic of interest in understanding and adapting Situationally-Induced Impairments and Disabilities. Designing for SIIDs is imperative because it will encourage designers to think about the entire spectrum (temporary/long- term/situational disabilities) that can help enhance and widen the usability for any user in any situational case.